\documentstyle[aps,epsf,prl,multicol]{revtex}

\begin{document}
\title{
Anomalous heat conduction and anomalous diffusion in one
dimensional systems}
\author{
Baowen Li$^a$ and Jiao Wang$^b$ }
\address{
$^a$Department of Physics, National University of Singapore,
117542 Singapore \\
$^b$ Temasek Laboratories, National University of Singapore, 119260 Singapore
}
\date{Submitted on 18 Dec. 2002, accepted for publication by Phys. Rev. Lett. on 14 June 2003}
\maketitle

\begin{abstract}
We establish a connection between anomalous heat conduction and
anomalous  diffusion in one dimensional systems. It is shown that
if the mean square of the displacement of the particle is
$\langle\Delta x^2\rangle =2Dt^{\alpha} (0<\alpha\le 2)$,  then
the thermal conductivity can be expressed in terms of the system
size $L$ as $\kappa = cL^{\beta}$ with $\beta=2-2/\alpha$. This
result predicts that a normal diffusion ($\alpha =1$) implies a
normal heat conduction obeying the Fourier law ($\beta=0$), a
superdiffusion ($\alpha>1$) implies an anomalous heat conduction
with a divergent thermal conductivity ($\beta>0$), and more
interestingly, a subdiffusion ($\alpha <1$) implies an anomalous
heat conduction with a convergent thermal conductivity
($\beta<0$), consequently, the system is a thermal insulator in
the thermodynamic limit. Existing numerical data support our
results.

\end{abstract}

\begin{multicols}{2}

Does heat conduction in one dimensional (1D) systems obey the
Fourier law? If it does, what are the necessary and sufficient
conditions? If it does not, then what is the reason and how the
thermal conductivity diverges/converges with the system size $L$?
These questions have attracted increasing attention in recent
years
\cite{Bon,Casati84,Prosen92,FPU,HLZFK,Hata,Tong,Prosen00,HLZPHI4,Li01,Gendelman,alonso,Aoki01,SG02,Li02,Li03a,Li03b,nanotubes,binarysphere,Grassberger,NR02}.
Although some progress have been achieved, many puzzles remain.
For example, in an attempt to establish a connection between heat
conduction and the underlying microscopic dynamics, there exist
some controversial examples. In the ding-a-ling model Casati et
al\cite{Casati84} show that as onset of global chaos the heat
conduction crosses over from an abnormal one to a normal one
obeying the Fourier law. It is thus concluded that the chaos is a
deciding factor. Later on, in order to show the exponential
instability is a necessary condition, Alonso et al \cite{alonso}
studied the heat conduction in a Lorentz gas channel, a quasi 1D
billiard with circular scatterers, and found that the heat
conduction obeys the Fourier law. However, the results from 1D
Ehrenfest gas channels \cite{Li02}, in which the Lyapunov
exponent is zero, show that the Fourier heat law might not have
any direct connection to the underlying dynamical chaos, because
the heat conduction can be normal and abnormal, depending on
whether or not the disorder is introduced.

Recently, a quasi 1D triangle billiard model, which consists of
two parallel lines of length $L$ at distance $d$ and a series of
triangular scatterers, has been introduced and
studied\cite{Li03a}. In this model, no particle can move between
the two reservoirs without suffering elastic collisions with the
triangles. Therefore this model is analogous to the Lorentz gas
channel studied in \cite{alonso} with triangles instead of discs
and the essential difference is that in the triangular model the
dynamical instability is linear and therefore the Lyapunov
exponent is zero. It is found that the motion inside the
irrational triangle channel (the internal angles are irrational
multiples of $\pi$) is diffusive and has a normal heat conduction.
Therefore deterministic diffusion and normal heat transport which
are usually associated to full hyperbolicity, can take place in
systems without exponential instability. Another example is the
Fermi-Pasta-Ulam (FPU) model\cite{FPU} which has non-zero Lyapunov
exponent, however the heat conduction in this model does not obey
the Fourier law.

The heat conduction in the rational triangle model (the internal
angles are rational multiples of $\pi$) and in the FPU model is
anomalous and does not obey the Fourier law, the thermal
conductivity $\kappa$ diverges with system size $L$ as $L^{\beta}$
with $\beta=0.22$ for the rational triangle model\cite{Li03a}, and
$0.34<\beta<0.44$ for the FPU model\cite{FPU}. Indeed, similar
divergent behavior has been observed in many 1D systems. For
example, in the binary hard sphere
model\cite{binarysphere,Grassberger}, $0.22<\beta <0.35$, in
single wall nanotubes $0.22<\beta < 0.37$ \cite{nanotubes}, and in
many classical lattices such as the harmonic lattice,
$\beta=1$\cite{Lebowitz70}, disordered harmonic lattice,
$\beta=1/2$ \cite{Lebowitzdisorder}, and the Frenkel-Kontorova
(FK) model under the condition of $T/K>>1$,
$0<\beta<1$\cite{SG02}, where $T$ is temperature, and $K$ is the
effective amplitude of a sinusoidal on-site potential.

Obviously, a universal value of $\beta$ does not exist, it differs
from model to model. Most recently, Narayan and
Ramaswamy\cite{NR02} show theoretically that in a 1D
momentum-conserving continuous system, the heat conduction is
anomalous, and the thermal conductivity diverges with system size
$L$ as $L^{1/3}$. Up to now, in all available numerical results
only the heat conduction in a (5,5) single wall
nanotubes\cite{nanotubes} shows an exponent ($\beta\approx 0.32$)
close to this 1/3\cite{Linotes}. Despite the conduction mechanism
is similar, a (10,10) single wall nanotube shows different
value\cite{nanotubes} for unknown reasons. The numerical results
from other models such as the FPU model, the harmonic model and
other billiards models deviate largely from this value for reasons
to be investigated.

On the other hand, even if the momentum conservation breaks down,
the heat conduction can be anomalous such as that in the
Frenkel-Kontorova model\cite{SG02}. The question becomes how to
explain this anomalous heat conduction, in particular the value of
the exponent $\beta$ in the thermal conductivity. A general theory
is still lacking. The only existing theory is for the 1D harmonic
chain\cite{Lebowitz70}, in which the phonons transport along the
chain ballistically and the thermal conductivity, $\kappa$,
diverges as $L$, i.e. $\beta=1$.

In this Letter, we would like to find a microscopic origin of the
anomalous heat conduction observed in many 1D models. We shall not
restrict to any specific model. This should give us a more general
way to understand the heat conduction in 1D systems.

As is well known that, depending on the value of exponent $\alpha$
in the mean square of displacement of the particle, $\langle\Delta
x^2\rangle=2Dt^{\alpha}$ with $0<\alpha\le2$,  1D microscopic
motion can be classified into ballistic motion, $\alpha=2$,
superdiffusion, $1<\alpha<2$, normal diffusion, $\alpha=1$, and
subdiffusion $\alpha<1$. Ballistic transport is observed in the
harmonic lattice. Normal diffusion shows up in the FK model in a
certain parameter regime\cite{HLZFK}, the disordered FPU
model\cite{Li01}, the Lorentz gas channel\cite{alonso}, the
disordered Ehrenfest gas channel\cite{Li02}, the irrational
triangle channel\cite{Li03a}, and the alternative mass hard-core potential model\cite{Li03b}.
In some billiard models,
superdiffusion\cite{Li02,Li03a,Alonso02,liwang03,zaslavsky} and
subdiffusion\cite{Alonso02} are observed. Superdiffusion and
subdiffusion can be studied from the fractional Fokker-Planck
equation, for detailed theoretical investigation and discussion
about the anomalous diffusion, please refer to review
articles\cite{zaslavsky,anomalousdiffusion} and the references therein.

To establish a connection between the microscopic
process and the macroscopic heat conduction, let's consider a 1D
model of length $L$ whose two ends are put into contact with
thermal baths of temperature $T_L$ and $T_R$ for the left end and
the right end, respectively. Suppose the energy is transported by
energy carriers (they are phonons in lattices and particles in
billiard channels) from left heat bath to the right heat bath and
vice versa. If the mean square of displacement of the carrier,
with velocity $v$, inside the system can be described by,
\begin{equation}
\langle \Delta x^2\rangle = 2Dv^{\alpha}t^{\alpha}
\label{Diffusion}
\end{equation}
then the so-called ``mean first passage time" (MFPT)
is\cite{Gitterman00}

\begin{equation}
\langle t_{LR}\rangle= \frac{4\gamma}{\alpha \pi
v}\left(\frac{2L}{\pi\sqrt{D}}\right)^{\frac{2}{\alpha}}, \quad
\gamma= \sum_{n=0}^{\infty}
\frac{(-1)^n}{(2n+1)^{1+\frac{2}{\alpha}}}. \label{MFPT}
\end{equation}
Obviously, if the 1d system is isotropic, the MFPT for the carrier
travelling from the right to the left end $\langle t_{RL}\rangle$
is the same as  $\langle t_{LR}\rangle$.

If the heat bath is a stochastic kernel of Gaussian type, namely,
the probability distribution of velocities is $p(v,T)= 4\pi v^2
\exp(-v^2/2T)/(2\pi T)^{3/2}$, the MFPT becomes,

\begin{equation}
\langle t_{LR}\rangle = \frac{16T\gamma}{\alpha (2\pi
T)^{\frac{3}{2}}}\left( \frac{2L}{\pi
\sqrt{D}}\right)^{\frac{2}{\alpha}}.
 \label{MFPT2}
\end{equation}

We define the heat current as the energy exchange between two heat
baths in unit time. Thus the current induced by a carrier ($m=1$)
with velocity $v$ moves from left to right and comes back is:

\begin{equation}
j
=\frac{\int_0^{\infty}\frac{v^2}{2}\left(p(v,T_L)-p(v,T_R)\right)}{\langle
t_{LR}\rangle+\langle t_{RL}\rangle}=\frac{T_L-T_R}{2\langle
t_{LR}\rangle}, \label{flux}
\end{equation}
If the temperature difference between the two baths is sufficient
small so that $\nabla T = (T_R-T_L)/L$, then the thermal
conductivity, $\kappa=-Lj/\nabla T$, is

\begin{equation}
\kappa  =c L^{\beta},\qquad
 \beta=2-2/\alpha,
\label{conductivity}
\end{equation}
and the constant, $ c=
3\pi\sqrt{2\pi}\alpha(\pi\sqrt{D}/2)^{2/\alpha}\sqrt{T}/(32\gamma)$.

Eq. (\ref{conductivity}) is the central result of the
paper\cite{Linoteseq5}. It connects heat conduction and diffusion
quantitatively. The main conclusion is that an anomalous
diffusion indicates an anomalous heat conduction with a divergent
(convergent) thermal conductivity. More precisely, our result
tells us that: a ballistic motion means thermal conductivity
proportional to the system size $L$, a normal diffusion means a
normal heat conduction obeying the Fourier law, a superdiffusion
means a divergent thermal conductivity, a subdiffusion means a
zero thermal conductivity in the thermodynamic limit. In the
following, we compare our results with the existing
analytical and numerical results.

{\it A ballistic motion}, $\alpha=2$, leads to a divergent thermal
conductivity $\kappa \propto L$. The only existing analytical
result is heat conduction in a 1D harmonic lattice. It is known
that heat is transported by phonons in lattice model. Because
there are no resistance and umklapp process, the phonons transport
ballistically in harmonic lattice model, thus $\alpha =2$. From
our formula (\ref{conductivity}), the thermal conductivity in the
1D harmonic lattice diverges as $L^{\beta}$ with $\beta=1$, this
is exactly what was shown by Lebowitz et al\cite{Lebowitz70}
(``$\ast$" in Fig. 1).

{\it A normal diffusion}, $\alpha=1$, means that the thermal
conductivity is a size independent constant, $\beta=0$,
i.e. the heat conduction obeys the Fourier law. For example in the
1D Frenkel-Kontorova model\cite{HLZFK}, in a certain range of
parameter such as $T/K <<1$\cite{LiFKnotes}, the phonons transport
diffusively\cite{Tong}, thus the thermal conductivity is finite
and independent of the system size $L$. The disordered FPU model
also has a finite thermal conductivity due to the random walk like
scattering process in the chain\cite{Li01}. Other 1D models
showing normal diffusion and normal thermal conduction are: the 1D
Lorentz gas channel\cite{alonso}, the 1D disordered Ehrenfest gas
channel\cite{Li02}, the 1D irrational triangle
channel\cite{Li03a}, the alternative mass hard-core potential model\cite{Li03b}, 
and some 1D polygonal billiard channels with
certain rational triangles\cite{Alonso02}. ``$\star$" in Fig. 1
represents all models with normal diffusion.

\begin{figure}
\epsfxsize=8.cm \epsfbox{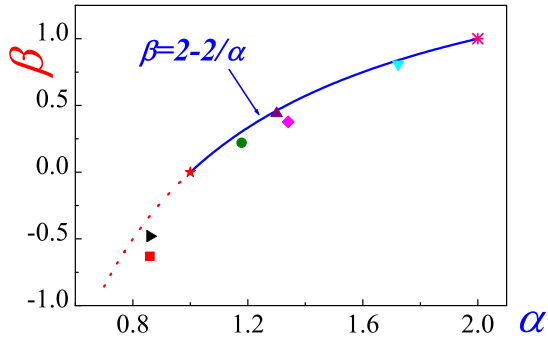}
\caption
{
The $\alpha-\beta$ plot. {\em Normal diffusion}:
$\star$ represents models with a normal diffusion and a normal
heat conduction, i.e. $\alpha=1$ and $\beta=0$, such as the
Lorentz gas channel[12], the Frenkel-Kontorova
model[5], the $\phi^4$ model[9], the
disordered FPU model[10], the disordered Ehrenfest gas
channel[15], the irrational triangle channel[16],
the alternative mass hard-core potential model[17],
and some rational polygonal channel[25] etc. 
{\em Ballistic motion}: $\ast$ represents the ballistic transport,
i.e. $\alpha=2$ and $\beta=1$, such as the 1D harmonic lattice
model. {\em Superdiffusion}: $\bigtriangledown$, 1D Ehrenfest gas
channel with right angle triangle scatterers[15];
$\bigcirc$, 1D channel with rational triangle
scatterers[16]; $\triangle$, polygonal billiard channel
with one irrational ($(\sqrt{5}-1)\pi/4$) and one rational
($\pi/3$) triangle; $\Diamond$, a 1D triangle-square
channel[26]. {\em Subdiffusion}:  the polygonal
billiard channel with one irrational angle ($(\sqrt{5}-1)\pi/4$)
and one rational angle ($\pi/4$)[25], $\Box$, from the channel length
$1 \le L \le 40$; $\triangleright$, from the channel of length
$ 40 \le L \le 80.$
}
\label{figure1}
\end{figure}

{\it A superdiffusion}, $1<\alpha<2$, implies an anomalous heat
conduction with a divergent thermal conductivity $L^{\beta}$. The
exponent $0<\beta<1$ differs from model to model. Here we take the
billiard models as our examples because they are very clean, and
both the diffusion and thermal conductivity in these models can be
calculated very accurately. The first example is the 1D Ehrenfest
gas channel in which the scattering obstacles are isosceles right
triangles periodically post along the channel\cite{Li02}. In this
model one has $\alpha=1.672$. From our analytical result
(\ref{conductivity}), the thermal conductivity should diverges as
$L^{\beta}$ with $\beta=2-2/\alpha=0.804$ which agrees with the
result from simulation of heat conduction
$\beta=0.814$\cite{Li02} (``$\bigtriangledown$" in Fig. 1). The
second example is the 1D channel with triangles whose inter angles
are rational multiples of $\pi$\cite{Li03a}. This model shows a
superdifussion with $\alpha=1.178$. The divergent exponent of
thermal conductivity is $\beta=0.302$. This exponent is slightly
larger than the one obtained from thermal conductivity simulation
$\beta=0.22$ (``$\bigcirc$" in Fig. 1). This deviation is due to the
finite size effect in the heat conduction simulation.

{\it A subdiffusion}, $\alpha<1$, results in an anomalous heat
conduction with a convergent thermal conductivity, i.e. $\kappa
\propto L^{\beta}$, with $\beta<0$. This is an interesting result
implying that the system becomes a thermal insulator in the
thermodynamic limit $L\to \infty$. Although there are many
examples showing subdiffusion
\cite{porous,percolation,fractal,polymer,semicond}, a systematic
study on the heat conduction in such kind of systems is still
lacking. The only existing example is the heat transport in a
polygonal billiard which supports our conclusion\cite{Alonso02}.
Most recently, Alonso et al\cite{Alonso02} show that in a very
special configuration, $\alpha=0.86$, and the thermal conductivity
goes as $\kappa \sim L^{-0.63}$ (``$\Box$" in Fig. 1). As $L$ goes
to infinity, the thermal conductivity goes to zero. According to
our formula (\ref{conductivity}), if $\alpha=0.86$, $\beta =-0.33$
which is larger than the one obtained by Alonso et
al\cite{Alonso02}. This is not a surprise, because the channel
length in their study of thermal conductivity is too small ($L\le
40$). If the channel is longer, the value of $\beta$ will become
much more closer to our theoretical estimation ($\beta=-0.33$). To
demonstrate this, we extend the thermal conductivity simulation
from $L\in[1,40]$ used by Alonso et al \cite{Alonso02} to
$L\in[40,80]$, and find that $\beta=-0.48$ (``$\rhd$" in Fig. 1)
which is more closer to $\beta=-0.33$ than the one obtained by
Alonso et al. If $L \to \infty$, one can expect $\beta$ goes to
$-0.33$.

All numerical results are summarized and represented in Fig 1,
where we draw $\beta$ versus $\alpha$, and compared with 
Eq. (\ref{conductivity}). As is shown that, Eq. (\ref{conductivity}) is 
exact for both normal diffusion and the ballistic
motion. The agreement with most existing numerical data is good.
 However, discrepancies remain for some
models mainly due to the limited numerical simulations. The best
data close to curve $\beta=2-2/\alpha$ is the simulation from 1D
Ehrenfest gas channel\cite{Li02}. This is because the channel length used in
the simulation is the longest one ($L\sim 10^3$) among all the
available data.

In summary, we have established a connection between anomalous
heat conduction and anomalous diffusion in 1D systems. Our central
result Eq. (\ref{conductivity}) includes all possible cases
observed in different classes of 1D models, ranging from
subdiffsion, normal diffusion, and superdiffusion to ballistic
transport. Several conclusions can be drawn: (1) A normal
diffusion leads to a normal heat conduction obeying the Fourier
law. (2) A ballistic transport leads to an anomalous heat
conduction with a divergent thermal conductivity  $\kappa\propto
L$. (3) A superdiffusion leads to an anomalous heat conduction
with a divergent thermal conductivity in thermodynamic limit. (4)
More importantly, our result predicts that a subdiffusion system
will be a thermal insulator. Existing numerical data support our
results.

We should mention that the subdiffusion process has been observed
in many real physical systems such as highly ramified media in
porous systems\cite{porous}, percolation
clusters\cite{percolation}, exact fractals\cite{fractal}, the
motion of a bead in a polymer network\cite{polymer}, charge
carrier transport in amorphous semiconductors\cite{semicond}. Any
numerical simulation and/or real experimental measurement of
thermal conductivity in these systems will be very interesting and
will allow one to test the theory given in this Letter. More
importantly, it will have a wide application in designing novel
thermal devices.

\bigskip
We would like to thank D Alonso for providing us
reference\cite{Alonso02} before publication, and G Casati for using discussions. This work was
supported in part by Academic Research Fund of NUS. JW is supported by DSTA
Singapore under Project Agreement POD0001821. This paper has been reported at 
the international workshop and seminar "Microscopic Chaos and Transport in Many-Particle Systems", at the MPIPKS Dresden from August 5 to August 25, 2002.

\end{multicols}


\begin{thebibliography}{9}
\bibitem{Linotes}It should be noted that in the binary sphere
model, some authors\cite{Grassberger} claim that the divergent
exponent is close to Narayan and Ramaswamy's theoretical
prediction 1/3, while the others\cite{binarysphere} disagree. A
careful and thorough study is needed to clarify this disagreement.
\bibitem{Linoteseq5}Alternatively, Eq.(\ref{conductivity}) can be derived
from the fractional Fokker-Planck equation by replace time $t$ by
$t^{\alpha}$.
\bibitem{LiFKnotes}It is worth mentioning that heat conduction in
the Frenkel Kontorova model obeys the Fourier law only in an
intermediate regime of parameter $T/K$. It is clear that in the
two extrmes, $T/K <<1$ and $T/K>>1$, heat conduction violates the
Fourier law.  The reason is that, in the case of $T/K<<1$ the
sinusoidal on-site potential can be approximated by an harmonic
on-site potential, in this case a temperature gradient cannot be
set up, the heat conduction is much like the case of an harmonic
chain. In the case of $T/K>>1$, the thermal fluctuation is so
large that the on-site potential is smeared out, namely, the
on-site potential will not be an effective barrier for the phonon
transport. Thus the role of on-site potential can be negelected.
For more quatitative analysis of the crossover from a normal to an
anomalous heat conduction in the Frenkel-Kontorova model, please
see Ref. \cite{SG02}.
\end{thebibliography}
\end{document}